\documentstyle[amsfonts,amssymb,amsmath,graphicx]{article}

\begin{document}

\date{}
\title{SCATTERING IN QUANTUM TUBES}
\author{B\"{O}RJE NILSSON\\ {\small School of Mathematics and Systems Engineering, V\"{a}xj\"{o}
University,} \\ {\small SE-351 95 V\"{A}XJ\"{O}, Sweden} \\ {\small E-mail: borje.nilsson@msi.vxu.se}}%
\maketitle
\begin{abstract} It is possible to fabricate mesoscopic
structures where at least one of the dimensions is of the order of
de Broglie wavelength for cold electrons. By using semiconductors,
composed of more than one material combined with a metal
slip-gate, two-dimensional quantum tubes may be built. We present
a method for predicting the transmission of low-temperature
electrons in such a tube. This problem is mathematically related
to the transmission of acoustic or electromagnetic waves in a
two-dimensional duct. The tube is asymptotically straight with a
constant cross-section. Propagation properties for complicated
tubes can be synthesised from corresponding results for more
simple tubes by the so-called Building Block Method. Conformal
mapping techniques are then applied to transform the simple tube
with curvature and varying cross-section to a straight, constant
cross-section, tube with variable refractive index. Stable
formulations for the scattering operators in terms of ordinary
differential equations are formulated by wave splitting using an
invariant imbedding technique. The mathematical framework is also
generalised to handle tubes with edges, which are of large
technical interest. The numerical method consists of using a
standard MATLAB ordinary differential equation solver for the
truncated reflection and transmission matrices in a Fourier sine
basis. It is proved that the numerical scheme converges with
increasing truncation.
\end{abstract}

\section{Introduction}

In the search for faster computers critical parts are becoming smaller. Today,
it is possible to build mesoscopic structures where some dimensions are of the
order of the de Broglie wavelength for cold electrons. Often the electron
motion is confined to two dimensions. Consequently, it may be necessary, at
least for some computer parts, to include quantum effects in the design process.

A large number of studies, devoted to such quantum effects, have
been carried out in recent years and a review is given by Londegan
et al~\cite{lo}. Many investigations aim at understanding the
physical properties of a particular quantum tube rather than
developing reliable mathematical and numerical methods that can be
used in a more general context. The research has given valuable
knowledge on the physical behaviour but also reports on the
limitations of the methods used. For instance, Lin \& Jaffe
\cite{li} report that a straightforward matching at the boundary
of a circular bend does not converge, demonstrating the numerical
problems with such a method. An illposedness is present in quantum
tube scattering and some type of regularisation is therefore
required to avoid large errors. Often, the tubes have sharp
corners to facilitate manufacturing but also to enhance quantum
effects. The presence of corners with attached singularities
requires special treatment.

Scattering of electrons in quantum tubes, see figure 1, is
theorywise related to the scattering of acoustic and
electromagnetic waves in ducts. Nilsson \cite{ni01} treats a
general method for the acoustic transmission in curved ducts with
varying cross-sections. Wellposedness, i.e. stability, is achieved
in an asymptotic sense. The mathematical framework guarantees
consistent results and allows for sharp corners and a proof for
numerical convergence is given. We set out to present a quantum
version of the results of Nilsson \cite{ni01}. In this way the
problems reported on convergence \cite{li} and on inconsistent
mathematical results would be resolved.

\begin{figure}
[h]
\begin{center}
\includegraphics[
scale=0.4, trim = 0 0 0 40]%
{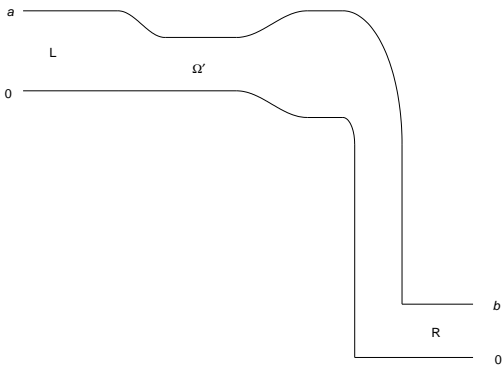}%
\caption{Two-dimensional quantum tube}%
\end{center}
\end{figure}

The paper is organised as follows. An introduction to scattering
in quantum tubes is given in section 2 and a mathematical model is
formulated in section 3. The Building block Method which is a
systematic method to analyse complicated tubes in terms of results
for simple tubes is also briefly described. Then in section 4 the
scattering problem for the curved tube with varying cross-section
and constant potential is reformulated to a scattering problem for
a straight tube with a varying refractive index. The solution to
this problem is presented in section 5 and a discussion on
numerical methods are also given.

\section{Tubes in quantum heterostructures}

A schematic view of a quantum heterostructure is shown in figure 2
following Wu et al. \cite{wuw} Electrons are emitted from the
n-type doped AlGaAs layer, migrate into the GaAs layer and stay
close to the boundary to the AlGaAs layer. In this way a very
narrow layer of electrons which are free to move in a plane is
formed. Nearly all the electrons in this two-dimensional gas are
in the same quantum state. By applying a negative potential on the
metal electrodes on the top of the heterostructure in figure 1,
the electrons are banished from the region below the electrodes.
For relatively low voltages, the effective potential in the tube
for one electron is close to the square-well potential. \cite{lo}
As a consequence the electrons in the two-dimensional gas are
further restricted to a tube that in form is a mirror picture of
the gap between the two electrodes. This quantum tube links the
electrons between the two two-dimensional gases on both sides of
the strip formed by the electrodes.

\begin{figure}
[h]
\begin{center}
\includegraphics[
scale=0.50,trim= 0 0 0 20
]%
{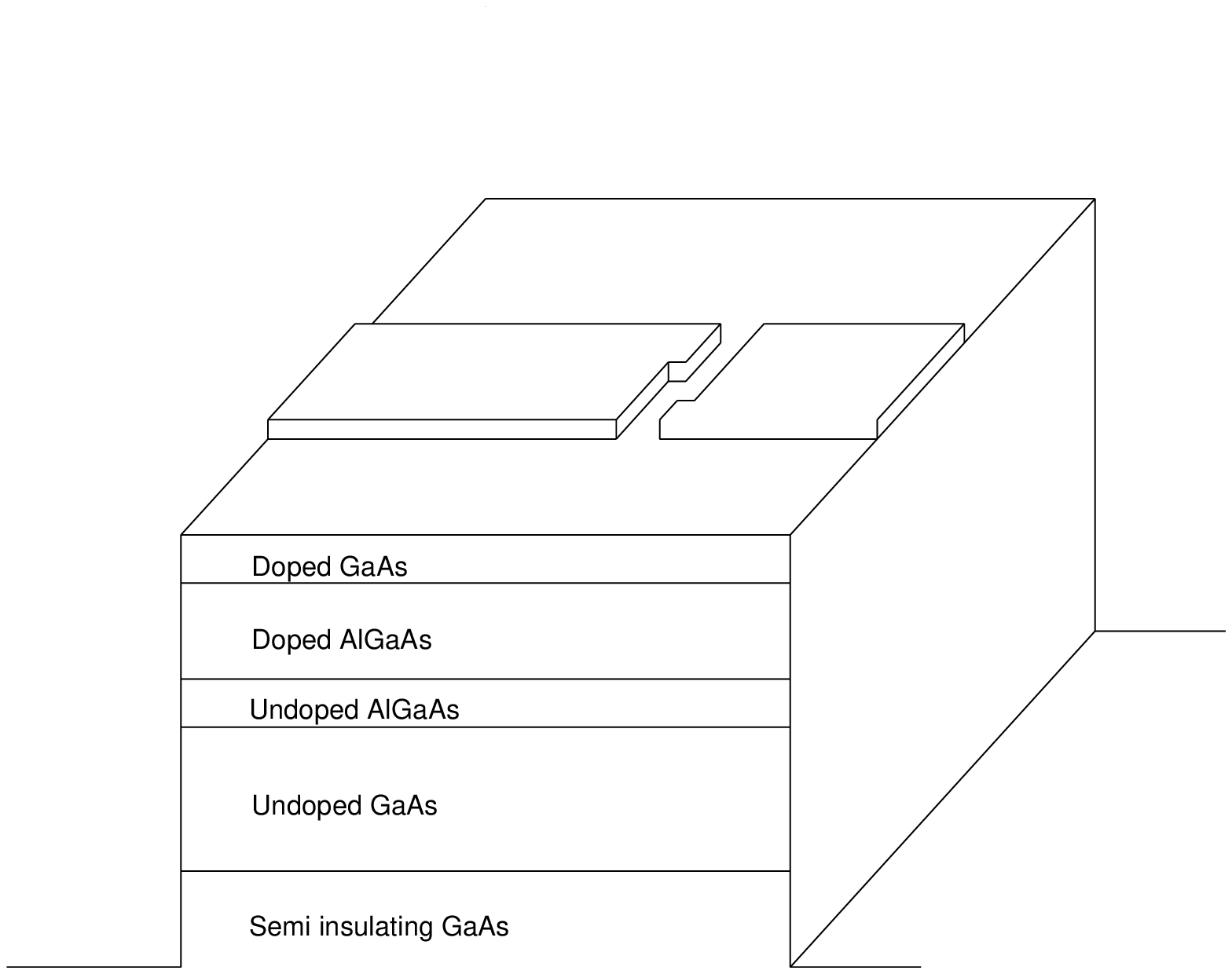}%
\caption{Schematic picture of heterostructure and split-gate structure.}%
\end{center}
\end{figure}

\section{Mathematical model}

Consider a two-dimensional tube with interior $\Omega^{\prime}$
according to figure 1. The boundary $\Gamma^{\prime}$ consists of
two continuous curves, $\Gamma_{+}^{\prime}$ and
$\Gamma_{-}^{\prime}$, which are piecewise C$^{2}$. The upper
boundary $\Gamma_{+}^{\prime}$ can be continuously deformed to
$\Gamma_{-}^{\prime}$ within $\Omega^{\prime}$. Outside a bounded
region the duct is straight with constant widths $a$ and $b,$
respectively. These terminating ducts are called the left and the
right terminating duct or L and R for short. We use stationary
scattering theory for one electron in an effective potential, with
time dependence exp$(-iEt/\hbar)$, assuming that the wave function
$\psi$ satisfies the time-independent Schr\"{o}dinger equation
$\triangle\psi +k^{2}\psi=0$ in $\Omega^{\prime},$where
$k^{2}=2m^{\ast}E/\hbar$ and $m^{\ast}$ is the effective mass
\cite{da}. Usually $k^{2}$ is called energy. The effective
potential is assumed to be a square well meaning that $\left.
\psi\right| _{\Gamma^{\prime}}=0.$

In a tube with constant cross-section the harmonic wavefunction $\psi$ can be
uniquely decomposed in leftgoing and rightgoing parts by $\psi=\psi^{+}%
+\psi^{-}$. Super indices $"+"$ and $"-"$ indicate rightgoing or plus and
leftgoing or minus waves respectively. Let $\psi_{in}^{+}$ and $\psi_{in}^{-}$
be known incoming waves in the terminating ducts. $\psi_{in}^{+}$ is present
in the left and $\psi_{in}^{-}$ in the right one. Let us write
\begin{equation}
\left\{
\begin{array}
[c]{c}%
\psi=\psi_{in}^{+}+R^{+}\psi_{in}^{+}+T^{-}\psi_{in}^{-}\text{ in L}\\
\psi=\psi_{in}^{-}+R^{-}\psi_{in}^{-}+T^{+}\psi_{in}^{+}\text{ in R}%
\end{array}
\right.  ,\label{1}%
\end{equation}
where for example the last two terms in (\ref{1}a) are minus waves and the
equation defines the left reflection mapping $R^{+}$ that maps the incoming
wave to an outgoing one in L. The scattering problem consists of finding the
mappings $R^{+},T^{-},R^{-}$ and $T^{+}$ as functions of energy for a given
duct. In summary we have
\begin{equation}
\left\{
\begin{array}
[c]{c}%
\triangle\psi+k^{2}\psi=0\text{ in }\Omega^{\prime}\\
\left.  \psi\right|  _{\Gamma^{\prime}}=0\\
\psi^{+}=\psi_{in}^{+}\text{ in L}\\
\psi^{-}=\psi_{in}^{-}\text{ in R}%
\end{array}
\right.  .\label{2}%
\end{equation}

There is always a solution to (\ref{2}), and except for a discrete
number of eigenenergies $k^{2}=k_{i}^{2},i=1,2,3,...,$ the
solution is unique. \cite{ce} When $k^{2}=k_{j}^{2},$ an
eigenenergy, there exists a solution without incoming but with
outgoing waves.

The use of the Building Block Method \cite{ni4} or transfer matrix
formalism \cite{wus} is very efficient for the solution of
scattering problems. In this method a tube with a complicated
geometry is divided into two parts usually where the tube is
straight. These two parts are converted to the type shown in
figure 1 by extending the terminating tubes to infinity. A sub
tube for the tube shown in figure 1 originates from the left part
and is depicted in figure 3. The Building Block Method gives a
procedure for calculating the mappings $R^{+},$ $T^{-},$ $R^{-},$
and $T^{+}$ for the entire tube in terms of the corresponding
scattering properties for the sub tubes. This procedure can be
repeated to get several sub tubes. Rather than using a general
numerical package for conformal mappings we have for the
calculations in this paper employed the Schwarz-Christoffel
mapping for a duct with corners and rounding the corners using the
methods of Henrici \cite{he}. Required analytic integrations are
performed in MATHEMATICA.

\begin{figure}
[h]
\begin{center}
\includegraphics[
scale=0.5,trim= -100 0 0 0
]%
{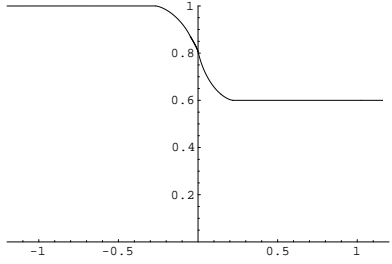}%
\caption{Sub-tube with interior $\Omega^{\prime}$ and upper
boundary
$\Gamma_{+}^{\prime}$and lower boundary $\Gamma_{-}^{\prime}.$ $b/a=0.6.$}%
\end{center}
\end{figure}

We recall the standard duct theory \cite{ce} in a form that illustrates the
illposedness of the problem and we have
\begin{equation}
\psi=\psi^{+}+\psi^{-}=\sum_{n=1}^{\infty}A_{n}^{+}e^{i\alpha_{n}x}\varphi
_{n}(y)+\sum_{n=1}^{\infty}A_{n}^{-}e^{-i\alpha_{n}x}\varphi_{n}(y),\label{3}%
\end{equation}
with $\varphi_{n}(y)=$ sin$(n\pi y/a)$ and $\alpha_{n}=\sqrt{k^{2}-n^{2}%
\pi^{2}/a^{2}},$ Im $\alpha_{n}\geq0.$ It is convenient to define the operator
$B_{0}$ by
\begin{equation}
\left\{
\begin{array}
[c]{c}%
B_{0}f=\sum_{n=1}^{\infty}\alpha_{n}f_{n}\varphi_{n,}\\
f(y)=\sum_{n=1}^{\infty}\alpha_{n}f_{n}\varphi_{n}(y)
\end{array}
\right.  .\label{4}%
\end{equation}
We find that $B_{0}^{2}=\partial_{x}^{2}+k^{2}$ and $\partial_{x}\psi^{\pm
}=\pm iB_{0}\psi^{\pm}.$ The initial value problem,
\begin{equation}
\left\{
\begin{array}
[c]{c}%
\partial_{x}\psi^{+}(x)=iB_{0}\psi^{+}(x),\\
\psi^{+}(0)=\psi_{0},
\end{array}
\right. \label{8}%
\end{equation}
is illposed for $x<0,$ but not for $x>0$. If an attenuated plus
wave is marched to the left an exponential growth is found. To
avoid the illposedness, $\psi$ is decomposed and the plus waves
are calculated by marching to the right and minus waves in the
opposite direction.

\section{Reformulated scattering problem}

To be able to use powerful spectral methods it is advantageous to transform
the tube to a flat boundary. It is enough, according to the Building Block
Method, to consider the scattering in the sub tubes and we restrict ourselves
to the first part as shown in figure 3. One way of transforming the tube is to
use a conformal mapping $w(\zeta)$ transforming the interior $\Omega^{\prime}$
of the tube with variable cross-section in the $\zeta=x+iy$ plane (figure 3)
to the interior $\Omega$ of a straight tube with constant cross-section in the
$w=u+iv$ plane. The straight tube is described by $-\infty<u<\infty,$ $0<v<a.$

\begin{figure}
[h]
\begin{center}
\includegraphics[
scale=0.6
]%
{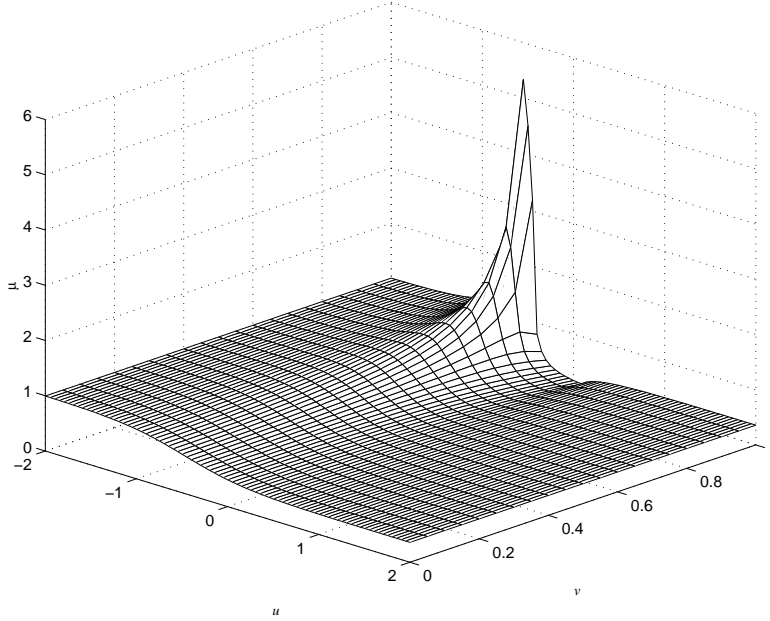}%
\caption{$\mu(u,v)$ in the straight duct. Parameters as in figure
3. $\mu
^{-1}$ is the refractive index.}%
\end{center}
\end{figure}

Introducing $\phi(u,v)=\psi(x,y)$ we get
\begin{equation}
\left\{
\begin{array}
[c]{c}%
\partial_{u}^{2}\phi+B^{2}(u)\phi=0\text{ in }\Omega\\
\phi(u,0)=\phi(u,a)=0,u\in\mathbb{R}%
\end{array}
\right.  ,\label{9}%
\end{equation}
with $B^{2}(u)=\partial_{v}^{2}+k^{2}\mu(u,v)$ and
$\mu=|d\zeta/dw|^{2}$. $\mu(u,v)^{-1}$ can be denoted as a
refractive index for the straight tube. In figure 4, $\mu$ related
to the simple tube in figure 3 is depicted. The factor $\mu(u,v)$
is asymptotically constant at both ends of the tube or more
precisely
$\mu(u,v)=\mu_{\pm}+$O$(e^{\mp|cu|}),u\rightarrow\pm\infty$ with
$\mu_{-}=1$ and $\mu_{+}=(b/a)^{2}$.

We use a first order description and rewrite (\ref{9}a) as%

\begin{equation}
\partial_{u}\left(
\begin{array}
[c]{c}%
\phi\\
\partial_{u}\phi
\end{array}
\right)  =\left(
\begin{array}
[c]{cc}%
0 & 1\\
-B^{2} & 0
\end{array}
\right)  \left(
\begin{array}
[c]{c}%
\phi\\
\partial_{u}\phi
\end{array}
\right)  .\label{13}%
\end{equation}
To avoid illposedness the decomposition $\phi=\phi^{+}+\phi^{-}$ is introduced
which must be identical to the corresponding decomposition (\ref{3}) in
regions where $\mu$ is a constant. The new state variables $(\phi^{+},\phi
^{-})$ are introduced via the linear relation
\begin{equation}
\left(
\begin{array}
[c]{c}%
\phi\\
\partial_{u}\phi
\end{array}
\right)  =\left(
\begin{array}
[c]{cc}%
1 & 1\\
iC & -iC
\end{array}
\right)  \left(
\begin{array}
[c]{c}%
\phi^{+}\\
\phi^{-}%
\end{array}
\right)  .\label{14}%
\end{equation}
Solving (\ref{14}) for $\phi^{+}$and $\phi^{-}$ and taking the $u$-derivative
and using (\ref{13}) we find that
\begin{equation}
\partial_{u}\left(
\begin{array}
[c]{c}%
\phi^{+}\\
\phi^{-}%
\end{array}
\right)  =\left(
\begin{array}
[c]{cc}%
\alpha & \beta\\
\gamma & \delta
\end{array}
\right)  \left(
\begin{array}
[c]{c}%
\phi^{+}\\
\phi^{-}%
\end{array}
\right)  ,\label{15}%
\end{equation}
where
\begin{equation}
\left\{
\begin{array}
[c]{c}%
\alpha=\frac{1}{2}[(\partial_{u}C^{-1})C+iC^{-1}B^{2}+iC]\\
\beta=\frac{1}{2}\left[  -(\partial_{u}C^{-1})C+iC^{-1}B^{2}-iC\right] \\
\gamma=\frac{1}{2}\left[  -(\partial_{u}C^{-1})C-iC^{-1}B^{2}+iC\right] \\
\delta=\frac{1}{2}\left[  (\partial_{u}C^{-1})C-iC^{-1}B^{2}-iC\right]
\end{array}
\right.  .\label{16}%
\end{equation}
To generalize the concept of transmission operators we make them
$u$-dependent, using a similar notation as Fishman \cite{fi}:
\begin{equation}
\left(
\begin{array}
[c]{c}%
\phi^{+}(u_{2})\\
\phi^{-}(u_{1})
\end{array}
\right)  =\left(
\begin{array}
[c]{cc}%
T^{+}(u_{2},u_{1}) & R^{-}(u_{1},u_{2})\\
R^{+}(u_{2},u_{1}) & T^{-}(u_{1},u_{2})
\end{array}
\right)  \left(
\begin{array}
[c]{c}%
\phi^{+}(u_{1})\\
\phi^{-}(u_{2})
\end{array}
\right)  ,\label{17}%
\end{equation}
assuming that $u_{1}\leq u_{2,}$ and suppressing the explicit $v$-dependence.
It is assumed for (\ref{17}) that the scattering problem has a unique solution
or that homogenous solutions are removed. A homogenous solution is usually
called a bound state.

Next we find a differential equation for the scattering operators $T^{+}%
(u_{2},u_{1}),$ $R^{-}(u_{1},u_{2}),$ $R^{+}(u_{2},u_{1})$ and $T^{-}%
(u_{1},u_{2})$ in (\ref{17}) using the invariant imbedding technique
\cite{be}$^{,}$ \cite{fi}. It is required that the incoming wave from the
right, $\phi^{-}(u_{2}),$ is vanishing. Then put $u_{1}=u,$ find $\partial
_{u}\phi^{-}(u)$ from (\ref{17}), use (\ref{17}) once more to obtain
\begin{equation}
\partial_{u}R^{+}(u_{2},u)=\gamma+\delta R^{+}(u_{2},u)-R^{+}(u_{2}%
,u)\alpha-R^{+}(u_{2},u)\beta R^{+}(u_{2},u),\label{18}%
\end{equation}
In a similar manner we get
\begin{equation}
\partial_{u}T^{+}(u_{2},u)=-T^{+}(u_{2},u)\alpha-T^{+}(u_{2},u)\beta
R^{+}(u_{2},u).\label{19}%
\end{equation}
The stability properties of (\ref{18}) and (\ref{19}) are of central
importance. In the flat regions where $B=B_{+}$ or $B_{-}$ we have $C=B$ and
$\partial_{u}C^{-1}=0$ implying that $\beta=\gamma=0$ and $\alpha=-\delta=iB.$
Similarly (\ref{18}) and (\ref{19}) reduce to $\partial_{u}X^{+}=-iBX^{+},$
$X^{+}=R^{+}$ or $T^{+},$ equations which are well-posed for marching to the
left. The initial values to accompany (\ref{18}) and (\ref{19}) are
$R^{+}(u_{2},u_{2})=0$ and $T^{+}(u_{2},u_{2})=I,$ where $I$ is the identity operator.

We choose $C=B_{-}+f(u)(B_{+}-B_{-})$ that is independent of $v$.
Here $f$ is increasing and smooth with
lim$_{u\rightarrow-\infty}f(u)=0,$ and
lim$_{u\rightarrow\infty}f(u)=1$.

\section{Solution of the scattering problem}

For the numerical solution of the scattering operator we expand $\phi$ in a
Fourier sine series and $\mu$ in a Fourier cosine series:
\begin{equation}
\left\{
\begin{array}
[c]{c}%
\phi(u,v)=\sum_{n=1}^{\infty}\phi_{n}(u)\varphi_{n}(v)\\
\mu(u,v)=\sum_{n=0}^{\infty}\mu_{n}(u)\xi_{n}(v)
\end{array}
\right.  ,\label{22}%
\end{equation}
where $\xi_{n}(v)=\cos(n\pi/a).$ Using the notation $\mathbf{\phi}=(\phi
_{0},\phi_{1},...)^{T}$ we find that
\begin{equation}
\frac{d^{2}\phi(u)}{du^{2}}+{\mathbf{B}}^{2}(u)\phi
(u)=0.\label{23}%
\end{equation}
The matrix elements of ${\mathbf{B}}^{2}(u)$ are given by
\begin{equation}
{\mathbf{B}}^{2}(u)_{nm}=\frac{k^{2}}{2}[-\mu_{m+n}(u)-\mu_{m-n}(u)-\mu_{m}%
+\mu_{n-m}(u)]-\frac{n^{2}\pi^{2}}{a^{2}}\delta_{nm}%
,\,n,m=0,1,2,...,\label{24}%
\end{equation}
and it is understood in (\ref{24}) that $\mu_{l}(u)=0$ for negative $l.$

For the tube in the physical $\zeta-$plane we require that locally both the
potential and the kinetic part of the energy are finite, that is both
$\int_{X}\left|  \psi\right|  ^{2}dxdy<\infty$ and $\int_{X}\left|  \nabla
\psi\right|  ^{2}dxdy<\infty$ for all finite regions $X$ inside the tube. We
say that $\psi$ belongs to the Sobolev space H$_{\text{loc}}^{1}$ meaning that
$\psi$ and its first derivatives are locally square integrable. Transformed to
the straight duct the local finite energy requirement means $\int_{U}\left|
\phi\right|  ^{2}\mu dudv<\infty$ and $\int_{U}\left|  \nabla\phi\right|
^{2}dudv<\infty$ for all finite regions $U$ inside the tube. For a smooth
boundary $\phi$ is more regular, and also the second derivatives of $\phi$ are
square integrable, that is $\phi\in$ H$_{\text{loc}}^{2}.$ It follows from the
theory of Grisvard \cite{gr} that also the second derivatives of $\phi$ are
square integrable, which means that $\phi\in$ H$_{\text{loc}}^{2}.$ According
to a graph theorem \cite{ta} $\phi\in$ H$_{\text{loc}}^{2}$ implies that
$\phi(u,\cdot)\in$ H$^{3/2}(0,a),$ meaning that up to 3/2 derivatives are
square integrable. To interpret this regularity with fractional derivatives we
define, following Taylor \cite{ta}, the function space
\begin{equation}
\text{D}_{s}=\left\{  f\in\text{L}^{2}(0,a);\sum_{n=0}^{\infty}|f_{n}%
|^{2}\left(  1+n^{2}\right)  ^{s}<\infty\right\}  ,s\geq0,\label{25}%
\end{equation}
with $f=\sum_{n=1}^{\infty}f_{n}\varphi_{n}$ and $f_{n}=(f,\varphi
_{n})/(\varphi_{n},\varphi_{n}).$ D$_{s}$ is a Hilbert space with the norm
\begin{equation}
||f||_{\text{D}_{s}}^{2}=(f,f)=\sum_{n=1}^{\infty}|f_{n}|^{2}\left(
1+n^{2}\right)  ^{s}.\label{26}%
\end{equation}
Taylor \cite{ta} shows that D$_{0}=$L$^{2}(0,a),$ D$_{1}=$H$_{0}^{1}(0,a),$
D$_{2}=$H$^{2}(0,a)\cap$H$_{0}^{1}(0,a)$ and that $\partial_{v}$D$_{s}=$
D$_{s-1},$ $s\geq1.$ In this terminology we have that for a smooth boundary
$\phi(u,\cdot)\in$ D$_{3/2}.$

The operator $\partial_{v}^{2}$ is self-adjoint on D$_{3/2}.$ Thus, we may
define $B_{\pm}$ by
\begin{equation}
B_{\pm}f=\sum_{n=1}^{\infty}\sqrt{k^{2}\mu_{\pm}-n^{2}\pi^{2}/a^{2}}%
f_{n}\varphi_{n},\label{28}%
\end{equation}
assuming that the branch Im $>$ 0 of the square root is taken. It
is clear that $T^{+},$ $R^{-},$ $R^{+}$ and $T^{-}$ are mappings
D$_{3/2}\rightarrow$D$_{3/2}$ and $B_{\pm}$: D$_{s}\rightarrow$
D$_{s-1},$ $s\geq1.$

For tubes with edges in the $\zeta-$duct things are a little more
complicated. With no restriction on the sharpness of the edges we
cannot improve that $\phi\in H_{\text{loc}}^{1}$ implying
$\phi(u,\cdot)\in$D$_{1/2}.$ Then, as an intermediate step in our
calculations $B_{\pm}\phi$ should be in the space D$_{-1/2}.$ Such
a derivative must of course be interpreted as a distribution.
However, the end result, i.e. scattered wave function belongs to
D$_{1/2}$. To generalise we define by
duality for positive $s$%
\[
\text{D}_{-s}=\left\{  g\text{
};\int_{0}^{a}f(v)g(v)dv<\infty\text{ for all
}f\in\text{D}_{s}\right\}  .
\]

Multiplication by$\sqrt{\mu}$ is an operator D$_{1/2}\rightarrow$ D$_{-1/2} $
and if $s\geq1/2$ we have the following mapping properties: $B_{\pm}:$ D$_{s}$
$\rightarrow$ D$_{s-1},\partial_{v}:$ D$_{s}$ $\rightarrow$ D$_{s-1},$ and
$T^{+},$ $R^{-},$ $R^{+}$ and $T^{-}$ are mappings D$_{s}\rightarrow$D$_{s}.$

The equations (\ref{18}-\ref{19}) can only in very special cases
be solved in a closed form. Therefore some type of numerical
scheme is used. Generally a numerical method cannot give uniform
convergence for the entire space $D_{s}.$ In a practical
application it is usually sufficient to know the effect of the
scattering matrices on the lowest eigenfunctions, the first
$N_{0}$ say. A practical method is therefore to truncate the
matrix representation of (\ref{18}) - (\ref{19}) to $N>>N_{0}$ and
solve the finite-dimensional ordinary differential equation with a
standard numerical routine. Nilsson \cite{ni01} proves that such a
procedure converges when $N\rightarrow\infty.$

Presently, numerical results are not available for the quantum tube
scattering. However, Nilsson \cite{ni01} presents results for the acoustic
case where the Neumann rather than the Dirichlet boundary condition applies.
He reports that for the lowest order reflection coefficient $N=1,$ i.e. a
scalar solution, is accurate up to $ka=1.5,$ $N=2$ gives a good and $N=5$
gives a perfect discription up to $ka=6.$ Energy conservation holds for all $N.$

\end{document}